\documentclass{article}
\usepackage{varioref}
\usepackage{amssymb}
\usepackage{color}
\begin{document}

\title{\bf  On the Stability of Einstein Static Universe \\ in \\
Doubly General Relativity Scenario }
\author{ M. Khodadi$^{1}$\thanks{
email: m.khodadi@stu.umz.ac.ir}\, , Y. Heydarzade$^{2}$\thanks{%
email: heydarzade@azaruniv.edu}\, , K. Nozari$^{1}$\thanks{email: knozari@umz.ac.ir}\, and F. Darabi $^{2,3}$\thanks{email: f.darabi@azaruniv.edu; Corresponding author}
\\{\small $^1$ \emph{Department of Physics, Faculty of Basic Sciences,
University of Mazandaran, P. O. Box 47416-95447, Babolsar, Iran}}
\\{\small $^2$\emph{Department of Physics, Azarbaijan Shahid Madani University, Tabriz, 53714-161, Iran} }\\
{\small $^3$ \emph{Research Institute for Astronomy and Astrophysics of Maragha (RIAAM), Maragha 55134-441, Iran}}}
\date{\today}
\maketitle
\begin{abstract}
By presenting a relation between average energy of the ensemble of probe photons and energy density of the Universe, in the context of {\it gravity's rainbow} or {\it doubly general relativity} scenario, we introduce a rainbow FRW Universe model. By analyzing the fixed points in flat FRW model modified by two well known rainbow functions, we find that the finite time singularity avoidance (i.e. Big-Bang) may still remain as a problem. Then, we follow the ``Emergent Universe'' scenario in which there is no beginning of time and consequently there is no Big-Bang singularity. Moreover, we study
the impact of a high energy quantum gravity modifications related to the gravity's rainbow on the stability conditions of an ``Einstein static Universe'' (ESU). We find that independent of a particular rainbow function, the positive energy condition  dictates a positive spatial curvature for the Universe. In fact, without raising a nonphysical energy condition in the quantum gravity regimes, we can address an agreement between gravity's rainbow scenario and basic assumption of modern version of ``Emergent Universe''. We show that in the absence and presence of an energy-dependent cosmological constant $\Lambda(\epsilon)$,  a stable Einstein static solution is available versus the homogeneous and linear scalar perturbations under the variety of obtained conditions. Also, we explore the stability of ESU against the vector and tensor perturbations.
\vspace{5mm}\noindent
\end{abstract}
\newpage
\section{Introduction}
In the framework of general relativity (GR),  the gravitational force is explained in terms
of the space-time curvature so that the field equations connect the space-time geometry to the
matter content. More technically, GR exhibits a Universe modeled by space-time with a mathematical
structure formed by four dimensional differentiable manifold \cite{Joshi}. It is commonly believed
that a unique mathematical framework to reconcile the quantum mechanics (QM) with GR is highly
dependent on our understanding of the space-time geometry. In other words, there is a comprehensive
agreement that the geometry of space-time is fundamentally explained by a quantum theory. It is
assumed that the Planck energy  $\epsilon_{pl}$  addresses a critical energy scale of transition from
classical GR to quantum gravity (QG), a theory which is supposed to challenge the most fundamental and
long standing issues of modern physics. Despite the lack of a complete theory of QG, it seems a semi-classical
or effective phenomenological approach of QG may guide us to disclose the mysterious nature of QG
\cite{Maggiore, Hossenfelder, Amelino, Smolin}. It is important to understand how to extract the testable
predictions from this fundamental theory. Interestingly, all semi-classical approaches offered so far unanimously
insist on the existence of a minimal measurement length in the nature,  known as the Planck length $l_{p}$.
Besides, it is believed that the modified mass-shell condition (or dispersion relations) arising from deviation
of Lorentz invariance in effective phenomenological approaches \cite{Amelino, Smolin}, may justify  some of the
phenomena taking place in astronomical
and cosmological scale, like the threshold anomalies of ultra high energy cosmic rays and TeV photons
\cite{Jacobson, Carroll}. It is noticeable that such Lorentz invariance violation is already predicted in the
context of other approaches to QG such as: non-commutative geometry \cite{Carroll},
spin network in Loop quantum gravity \cite{LQG}, and string field theory \cite{String}. Throughout the present
work, our attention specifically is focused on a semi-classical formalism knows as \emph{``Gravity's Rainbow"}
which has been designed by Magueijo and Smolin \cite{Magueijo}. In fact, it can be said that the rainbow
gravity is nothing but the doubly special relativity (DSR)  \cite{Amelino}   in the presence of
curved space-time  which is knows as the \emph{``Doubly General Relativity"}(DGR). In this formalism, there is
no single fixed space-time background, namely the space-time background appears as a geometric spectrum in terms
of the energy scale of the particle probe. Therefore, in the cosmological setting, we are dealing with the rainbow
modified metric by introducing a function in the Friedmann-Robertson-Walker (FWR) metric which depends on such variable
energy scale. Formation of the rainbow metric proposal involve various motivations such as the lack of a trivial definition
of the dual position space in DSR \cite{Magueijo}. However, this approach to QG is not problem free and  considerable
number of works have been focused on its various aspects \cite{them}.
Somehow,  the proposal of gravity's rainbow is similar to the idea of  \emph{``Running Coupling Constants''} in
particle physics and field theory, in the sense that at very small distances (high energy) the space-time
geometry is related to the energy scale of the  particle probe \cite{Visser}. It is
worth mentioning that, apart from the  effective approaches to QG like DSR and DGR, it has been
argued that if the  physical space-time of the standard GR theory is emergent, then it is expected that we face
with a radical picture of the Universe at fundamental quantum scale \cite{Visser}. The basic idea of
\emph{``Emergent Universe''} dates back to the works of Eddington in 1930 \cite{1930}, inspired by the proposal of
Einstein static Universe (ESU). In simple words, ESU suggests that the content of the Universe as a closed system
is made up of at least cosmological constant and normal matter. While ESU in the context of  GR is not stable versus
spatially homogeneous perturbations, it predicts that the Universe in past might have been emerged as an static
initial states \cite{Bag}. We know that the data obtained by cosmic microwave background (CMB) observations,
generally supports the inflationary scenario. Also, by  implementation of  some singularity theorems with the
geometrical assumptions that $K<0$ or $K=0$, it has been shown that despite the occurrence of inflation,
the Universe had a violent beginning in the past \cite{Borde}. Hence, one of the most important motivations
for studying QG is to avoid the singularity at the beginning of the Universe. Recovery of  ESU in the framework
of inflationary Universe by Ellise and Maartens, knows as modern version of Eddington emergent Universe, was in
this direction of research interest \cite{EM}. According to the modern standard cosmology,  the inflationary
expansion of the Universe has quickly eliminated any original spatial curvature. On the other hand, the CMB
observations implies  $\Omega_{0}\gtrsim1$ ($1.02\pm0.02$) for the total density parameter. This means that
the geometry of Universe is not exactly flat, rather it could have non-zero spatial curvature at the beginning,
resulting in negligible late-time effects. In this sense, the emergent Universe scenario allowed the existence of
a positive curvature Universe that emanates asymptotically as a static initial state known as ESU which afterwards
experienced the inflation. {The stable ESU addresses the existence of a fixed point around
which our Universe in the early times was eternally fluctuating. In the case of the Universe filled
with a massless scalar field $\phi$ with a suitable  inflaton potential,
these fluctuations had been disappeared and the stable state of Universe had been
turned  to inflationary phase. Within the context of
standard cosmology, the prerequisites of a one or further flat wings (of course with a little positive gradient)
in the inflaton potential, can make an emergent Universe.}  {This transmission
from a stable ESU to inflationary phase takes place around the different values of e-folds. For instance, in \cite{Bag}
by analyzing the spectrum of inflationary perturbations of some stable cosmological models in the context of GR,
it has been demonstrated that the transmission occurs in over 60 e-folds}. {Also, it should be
noted that  while these inflationary emergence
GR-based models respect to CMB constraints,  they suffer from a fine-tuning
problem. This problem can be resolved in the context of modified emergence models of GR. However, recently
in \cite{Aud} some arguments are presented which tries to show the emergent cosmological scenario can not really be past-eternal.
}As a prominent feature of emergent Universe scenario, one can point to the removal of
initial singularity and also the horizon problem. This is a strong motivation to study the ESU models along with
their stability conditions in the presence of high energy corrections of GR. For instance, one can point to the
works done in the context of  massive gravity \cite{massive}, Ho\v{r}ava-Lifshitz model of gravity \cite{Ho},
 braneworld scenarios \cite{brane}, induced matter theory \cite{Yaghoub}, loop quantum cosmology \cite{Loop}, $f(R)$, $f(T)$ and $f(G)$ gravity \cite{RTG}.
As discussed at the beginning of this section,
gravity's rainbow or DGR is also considered as an alternative of GR with high energy corrections. Besides, it has been shown
that in the study of FRW cosmology and  isotropic quantum cosmological perfect fluid model in rainbow gravity setup, some
conditions are derived which prevent the initial singularity \cite{Barun2}. Inspired by this introduction, our main goal in
this paper is to study the stability of ESU against the homogeneous scalar, vector and tensor perturbations in the presence
of rainbow's gravity setup as a QG modification of GR. {But before doing this, by following the phase space method we will analyze the status
of the Big-Bang singularity in the framework of a flat rainbow FRW Universe model.}  \\
The present paper is arranged as follows: In section 2, we have
derive rainbow modified Friedman
equations in details for a general non-flat universe. In section 3, by introducing two different forms
of rainbow functions depending on energy density, firstly in absence of cosmological constant,  we
obtain the stability conditions of ESU against the homogeneous linear scalar perturbations.
In the follow of this section, we pursue our goal in the presence of a cosmological constant related to
energy density via introducing a new rainbow function. In section 4, we analysis
the stability against the  vector and tensor perturbations. Finally in section
5, we give the summery and conclusions.

\section{Gravity's Rainbow Modified FRW Universe}
In this section,  firstly we have a quick review on the gravity's rainbow theory,
known also as DGR. Then,  with note to the importance of spatial curvature
in Emergent Universe scenario, we derive the modified
Friedman equations for general non-flat universes. In DSR,
the modified dispersion relations of a massive particle with mass $m$
reads as
\begin{equation}\label{e1-1}
\epsilon^{2}f^{2}(\epsilon/\epsilon_{p},\eta)-p^{2}g^{2}(\epsilon/\epsilon
_{p},\eta)=m^{2}\;,
\end{equation}
where $f(\epsilon/\epsilon_{p},\eta)$ and $g(\epsilon/\epsilon_{p},\eta)$ are
the well known energy dependent rainbow functions and $\eta$ is a dimensionless
parameter. In the low energy limit, the modified dispersion relation (\ref{e1-1})
reduces to the relativistic dispersion relations. So, the rainbow functions $f(\epsilon/
\epsilon_{p},\eta)$ and $g(\epsilon/\epsilon_{p},\eta)$  satisfy the following
conditions
\begin{equation}\label{e1-2}
\lim_{\epsilon/\epsilon_{p}\rightarrow0} f(\epsilon/\epsilon_{p},\eta)=1,\quad
\lim _{\epsilon/\epsilon_{p}\rightarrow0}g(\epsilon/\epsilon_{p},\eta)=1.
\end{equation}
Indeed, the condition (\ref{e1-2}) points to the correspondence principle. Based on the
discussion carried out in \cite{D}, for established position space in DSR, theories including
free fields must also lead to the plane wave solutions in flat space-time, despite satisfying
the modified dispersion
relations (\ref{e1-1}). For this reason, contraction between infinitesimal
displacement $dx^{a}$ and momentum $p_{a}$, must be linearly invariant, i.e
\begin{equation}\label{e1-3}
dx^{a}p_{a}=dt\epsilon+dx^{i}p_{i}\;.
\end{equation}
In fact, the linear contraction (\ref{e1-3}) guarantees the existence of the plane
wave solutions. The rainbow metric generally has the form of
\begin{equation}\label{e1-4}
ds^{2}=\frac{-1}{f^{2}(\epsilon/\epsilon_{p},\eta)}dt^{2}+\frac{1}{g^{2}
(\epsilon/\epsilon_{p},\eta)}dx^{2}\;.
\end{equation}
Using a one parameter family of energy momentum tensors, the gravity's rainbow modified
Einstein equations will be written as
\begin{equation}\label{e1-5}
G_{\alpha\beta}(\epsilon/\epsilon_{p})=8\pi G(\epsilon/\epsilon_{p})T_{\alpha\beta}
(\epsilon/\epsilon_{p})+g_{\alpha\beta}(\epsilon/\epsilon_{p})\Lambda(\epsilon/
\epsilon_{p})\;,
\end{equation}
so that $G(\epsilon/\epsilon_{p})=h_{1}(\epsilon/\epsilon_{p})G$ and $\Lambda(\epsilon/
\epsilon_{p})=h_{2}(\epsilon/\epsilon_{p})\Lambda$ where $h_{1}(\epsilon/\epsilon_{p})$
and $h_{2}(\epsilon/\epsilon_{p})$ are energy dependent rainbow functions. This means that
in DGR setup, the Newton’ian gravitational constant and the cosmological constants are energy
dependent such that in low energy limit, we recover $ G(\epsilon/\epsilon_{p})=G$ and $\Lambda
(\epsilon/\epsilon_{p})=\Lambda$. In order to achieve our main goal, in this section we will set
$\Lambda=0$ for simplicity, and begin with the following modified FRW metric of a homogeneous
and isotropic Universe
\begin{equation}\label{e2-1}
ds^{2}=-\frac{N(t)^{2}}{f^{2}(\epsilon/\epsilon_{p},\eta)}dt^{2}+\frac{1}{g^{2}
(\epsilon/\epsilon_{p},\eta)}a^{2}(t)h_{ij}dx^{i}dx^{j},
\end{equation}
where $h_{ij}$ represents the spatial part of the metric.
Commonly, it is assumed that the energy $\epsilon$ of particle probe for each measurement is constant
and independent of space-time coordinates. Nevertheless, such assumption for the measurements at early
Universe seems to be far from reality. Therefore, it is expected that the background metric of space-time
throughout its evolution is affected by the energy $\epsilon$ of particle probe. Then, it is reasonable
to consider the evolution of energy $\epsilon$ with the cosmological time, denoted as $\epsilon(t)$.
In this regard, we will derive the modified FRW equations. In doing so, the following ansatz is usually
suggested \cite{Ling}
\begin{equation}\label{e2-1a}
\eta=1,\quad g^{2}(\epsilon,\eta)=1\;,
\end{equation}
where $f$ has the form of $f(\epsilon/\epsilon_{p})$. Therefore, the modified Einstein field equation
can be written as
\begin{equation}\label{e2-2}
R_{\alpha\beta}=-8\pi G(\epsilon)S_{\alpha\beta}(\epsilon),
\end{equation}
where $R_{\alpha\beta}$ is Ricci tensor defined as follows
\begin{equation}\label{e2-3}
R_{\alpha\beta}=\frac{\partial\Gamma^{\lambda}_{\lambda\alpha}}{\partial x^{\beta}}-
\frac{\partial\Gamma^{\lambda}_{\alpha\beta}}{\partial x^{\lambda}}+\Gamma^{\lambda}_
{\alpha\mu}\Gamma^{\mu}_{\beta\lambda}-\Gamma^{\lambda}_{\alpha\beta}
\Gamma^{\mu}_{\lambda\mu},\quad \alpha,\beta=0,1,2,3,
\end{equation}
and $S_{\alpha\beta}(\epsilon)$ is written in terms of the energy momentum tensor $T_{\alpha\beta}
(\epsilon)$  as\begin{equation}\label{e2-4}
S_{\alpha\beta}(\epsilon)=T_{\alpha\beta}(\epsilon)-\frac{1}{2}g_{\alpha\beta}T^{\mu}_{\mu}
(\epsilon)\;.
\end{equation}
To continue our calculations, we need the non-zero components of the affine connection as \cite{Ling}
\begin{equation}\label{e2-5}
\Gamma^{0}_{00}=-\frac{\dot{f}}{f},\quad \Gamma^{0}_{ij}=f^{2}\dot{a}a\delta_{ij},
\quad \Gamma^{i}_{0j}=\delta^{i}_{j}\frac{\dot{a}}{a}\quad i,j=1,2,3\;.
\end{equation}
It is  seen that unlike the usual FRW metric, for modified FRW metric (\ref{e2-1}), the
components of the affine connection with two time indices remain non-zero. By putting
(\ref{e2-5}) into (\ref{e2-3}) and after a straightforward calculation, we obtain
\begin{equation}\label{e2-6}
R_{00}=3\frac{\ddot{a}}{a}+3\frac{\dot{a}}{a}\frac{\dot{f}}{f}\;,
\end{equation}
and
\begin{equation}\label{e2-7}
R_{ij}=\tilde{R}_{ij}-\left((a\ddot{a}+2\dot{a}^{2})f^{2}-f\dot{f}\dot{a}a\right)
\delta_{ij}\;,
\end{equation}
where $\tilde{R}_{ij}$ denotes the purely spatial Ricci tensor defined as
\begin{equation}\label{e2-8}
\tilde{R}_{ij}=\frac{\partial\Gamma^{n}_{ni}}{\partial x^{j}}-
\frac{\partial\Gamma^{n}_{ij}}{\partial x^{n}}+\Gamma^{m}_{in}
\Gamma^{m}_{jm}-\Gamma^{m}_{ij}
\Gamma^{n}_{nl}.
\end{equation}
The spatial components $\Gamma^{i}_{jk}$ of the four dimensional affine connection
are identical with those of  affine connection computed in three dimensions
from the spatial three-metric $h_{ij}$, i.e.
\begin{equation}\label{e2-8a}
\Gamma^{i}_{jk}=\frac{1}{2}h^{in}\left(\frac{\partial h_{jn}}{\partial x^{k}}
+\frac{\partial h_{kn}}{\partial x^{j}}-\frac{\partial h_{jk}}{\partial x^{n}}
\right)\equiv \tilde{\Gamma}^{i}_{jk}\;,
\end{equation}
where $\tilde{\Gamma}^{i}_{jk}$ denotes purely spatial affine connection and $h^{ij}$
is the inverse of the $3\times3$ matrix $h_{ij}$. Therefore, we have $\Gamma^{n}_{ij}=kx^{n}
h_{ij}$ which results in
\begin{equation}\label{e2-9}
\tilde{R}_{ij}=-2kh_{ij}\;.
\end{equation}
Here, $k$ has a geometrical interpretation. Indeed, it measures the spatial curvature with zero,
negative and positive $k$ values corresponding to flat, open and closed Universes respectively
\cite{Weinberg}. The Ricci tensor expression (\ref{e2-7}) can be rewritten as
\begin{equation}\label{e2-10}
R_{ij}=-\left((a\ddot{a}+2\dot{a}^{2})f^{2}-f\dot{f}\dot{a}a+2k\right)h_{ij}\;.
\end{equation}
Now, for  obtaining the components of $S_{\alpha \beta}$, we consider the perfect fluid with the
energy-momentum tensor as
\begin{equation}\label{e2-11}
T_{\alpha\beta}=(\rho+p)u_{\alpha}u_{\beta}+pg_{\alpha\beta}\;,
\end{equation}
so that $\rho$ and $p$ represent the energy density and the pressure, respectively.
Also, $u_{\alpha}$ is the velocity four vector  defined by $u_{\alpha}=(f^{-1},0,0,0)$,
with the norm of $g^{\alpha\beta} u_{\alpha}u_{\beta}=-1$. We should note that the diagonal
components of the modified FRW metric tensor $g_{\alpha\beta}$ is $\left(-f^{-2},h_{ii}\right)$
with the signature $(-,+,+,+)$ so that the components of spatial tensor $h_{ii}$ are independent
of the rainbow function $f$. Using the equation (\ref{e2-4}), one can decompose the time and spatial
components of $S_{\alpha\beta}$ tensor as follows
\begin{equation}\label{e2-12}
S_{tt}=T_{tt}+\frac{1}{2}g_{tt}T,\quad S_{ij}=T_{ij}-\frac{1}{2}h
_{ij}a^{2}T\;,
\end{equation}
where
\begin{equation}\label{e2-13}
T_{tt}=\rho f^{-2},\quad T_{ij}=h_{ij}a^{2}p\;,
\end{equation}
and
\begin{equation}\label{e2-14}
T=g^{\alpha\beta}T_{\alpha\beta}=(-\rho+3p)\;.
\end{equation}
By substituting expressions (\ref{e2-13}) and (\ref{e2-14}) into (\ref{e2-12}), we have
\begin{equation}\label{e2-15}
S_{tt}=\frac{1}{2f^{2}}(\rho+3p),\quad S_{ij}=\frac{1}{2}(\rho-p)a^{2}h_{ij}.
\end{equation}
Ultimately, the first and second rainbow modified Friedman equations for the metric
parameterized by the varying energy probe (\ref{e2-1}), takes the form of
\begin{equation}\label{e2-a}
(\frac{\dot{a}}{a})^{2}+\frac{k}{a^{2}}\frac{1}{f^{2}}=\frac{8\pi G\rho}{3}\frac{1}{f^{2}}
\;,
\end{equation}
\begin{equation}\label{e2-b}
\frac{\ddot{a}}{a}=-\frac{4\pi G\rho(1+3\omega)}{3}\frac{1}{f^{2}}-\frac{\dot{a}}{a}\frac{\dot{f}}{f}\;.
\end{equation}
Here, also for simplicity, it is assumed that gravitational constant $G$ is independent of energy
$\epsilon$ . It should be noted that the existence of $\dot{f}$ term
in the second rainbow modified Friedman equations, does not mean the explicit dependence of rainbow
function to the time i.e. $f(t)$, at all. As explained above, $f$ is an explicit function of the
energy of the test particles which are probing the geometry of space-time in early Universe. Given the
fact that $\epsilon$ can vary with respect to the evolution of  Universe, so $f$ can be an implicit function
of cosmic time and not explicit. We note that, the explicit dependence of $f$ to the time, may be leads to this
wrong result from equations (\ref{e2-a}) and (\ref{e2-b})  that they are nothing but the
usual Friedman equations of GR so that the rainbow function $f$ in this case is merely the gauge parameter
determining the choice of time. Hence, based on the gauge freedom, one may choose the gauge of $f=1$.
In contrast to this misconception, the gravity's rainbow theory is a high energy modified theory of GR in which according to
\emph{correspondence principle,} for the case of low energy limit i.e. $\frac{\epsilon}{\epsilon_{pl}}\rightarrow0$
($f\rightarrow1$), modified Friedman equations (\ref{e2-a}) and (\ref{e2-b}) reduce to the standard equations for
the FRW universe. {Also, we mention that due
to the existence of the rainbow
functions, $t$ may play the role of proper time based on the gauge choice $ N=\frac{f_{1}(\varepsilon)}{f_2^{3\omega}
(\varepsilon)}a^{3\omega}$ \cite{Barun2}. As expected,  in the limit of GR, $t$ represents the proper time indicated by the choice of
$N=a^{3\omega}$}. Also, by combining the modified Friedman equations (\ref{e2-a}) and (\ref{e2-b}), we get the following
energy conservation equation
\begin{equation}\label{e2-c}
\dot{\rho}+3\frac{\dot{a}}{a}\rho(1+\omega)=0\;.
\end{equation}
It can be mentioned that in framework of  DGR, the form of equation of state (EOS) $p=\omega\rho$, remains unchanged for massless
prob particles such as photons while for the massive ones, this equation of state will be modified, see \cite{ Alexander, Magueijo}
for more discussion.

\section{Einstein Static Universe and Scalar Perturbations}
\subsection{ In the Absence of Cosmological Constant $\Lambda$}
In this section, we plan to apply the linear homogeneous scalar perturbations
in the vicinity of the Einstein static Universe and explore its stability
against these perturbations. To achieve our goal, we need to fix the rainbow
function $f$. To this end, using dispersion relation offered in \cite{MS},
we introduce rainbow functions as follows
\begin{equation}\label{e3-1}
f(\epsilon)=(1-\frac{\epsilon}{\epsilon_{Pl}})^{-1}\;,
\end{equation}
where via suggesting the average energy $\bar{\epsilon}=\frac{4c}{3}\rho^{\frac{1}{4}}$
($c$ is some constant) \cite{Adel}, this rainbow function will be rewritten as
\begin{equation}\label{e3-2}
f(\rho)=(1-\frac{4}{3}\xi \rho^{\frac{1}{4}})^{-1}\;,
\end{equation}
where $\xi=\frac{c}{\epsilon_{Pl}}$. The authors of \cite{Adel}, to reach
the above relation considered a large ensemble of prob photons which are in thermal equilibrium. This assumption
is reasonable since based on the standard model of cosmology, early universe passed a radiation
dominated era with $p=\frac{1}{3}\rho$.  One  may ask the question that why the identification $\epsilon\sim \bar{\epsilon}$ is used to obtain the rainbow function (\ref{e3-2})?. The answer is that, we are deal with
the energy of prob particle in the rainbow metric as the statistical mean value of all prob photons
in radiation domination. Indeed, we deal with the average effect of
photon particles in radiation dominated era and no with a special elected photon from the radiation \cite{Ling}.
It is worthwhile to remind that the form of $\bar{\epsilon}$
in terms of $\rho$ is independent of the modified dispersion relation picked for a specific
model \cite{Barun1}. { We also mention that the above rainbow function
is valuable from the theoretical viewpoint  so that in \cite{MS} it is shown that in the absence of the varying speed of light (VSL)
proposal, it can removes the horizon problem in early Universe. Moreover, it
is seen that  when $\rho$
or $\bar{\epsilon}$ get their largest values, then $f(\rho)$ or $f(\bar{\epsilon})$ becomes infinite which causes the time-like
component of the rainbow metric vanishing. This means that the FRW metric which is modified with rainbow function (\ref{e3-2})
becomes degenerate at that time and has no inverse. Indeed, a degenerate metric addresses the existence of other distinct possibility for lightlike
dimension   (other than timelike and spacelike dimensions). Also,
remember that in the Palatini formulation of standard GR,  degenerate metric also
appears. One of the consequences the degenerate metric is that the curvature remains bounded and the topology of space-time can change. Overall, it is believed
that singularities arisen from degenerate metric have a milder manner than other type of singularities and seem appropriate for a QG proposal \cite{Louko}}.
Here, it is necessary to
review the finite time singularity issue, including the Big-Bang singularity, in the presence of
rainbow function (\ref{e3-2}). Indeed, we want to investigate whether the presence of rainbow
function (\ref{e3-2}) in this system will lead to resolve the singularity problem. Inspired by the
idea proposed in \cite{Strogatz, Adel}, we want to resolve this problem by finding an upper bound
on the density of energy $\rho$. In other words, according to \cite{Strogatz, Adel} the finite time
singularity issue will be eliminated via the presentation of a fixed point as $\rho_{f}$ for energy
density $\rho$ which is reached at an infinite time. {More exactly,
the author of \cite{Adel},  by following
the terminology used for stability analyzing the dynamical systems in \cite{Strogatz}, demonstrated that for
any first order system as $\dot{\rho}=O(\rho)$, the finite time singularity will be solved through the
fulfillment of either of the following conditions:
1) the function $O(\rho)$ be a continuous and differentiable on a range enclosed by zeroes of function
$O(\rho)$. 2) asymptotically, function $O(\rho)$ growths like  a linear function as $K(\rho)$ or slower
than it, i.e $K(\rho)\geq O(\rho)$. Therefore,  for cancelation of the finite time singularity, it is sufficient that one of these conditions is satisfied. We begin our analysis in this way by substituting
the rainbow function (\ref{e3-2}) into energy conservation equation (\ref{e2-c}) and modified first
Friedmann equation (\ref{e2-a}) (for case $k=0$)
to obtain the ordinary differential equation (ODE) as $\dot{\rho}=O(\rho)$
where
\begin{equation}\label{e3-3}
O(\rho)=-4\rho(1-\xi \rho^{\frac{1}{4}})\left
(\frac{8\pi G}{3}\rho\right)^{1/2}\;.
\end{equation}
The advantage of the dynamical system analysis method followed in \cite{Strogatz, Adel} is that by having fixed
points and regarding the asymptotic behavior of $O(\rho)$, one is able to predict the demeanor of
the system without need to
have a detailed form of the solutions. Hence, by setting $\omega=\frac{1}{3}$ (since our attention is on radiation
dominated state in order to study the initial singularity), equation (\ref{e3-3}) results in the following two fixed points
\begin{equation}\label{e3-4}
\rho_{f_{1}}=0, \quad \rho_{f_{2}}=(\frac{\epsilon_{Pl}}{c})^{4}\;.
\end{equation}
However, to understand the qualitative manner of a solution, we should know that how long it takes to get
a fixed point. By a straightforward calculation, one can show that the time required
to reach these fixed points, can be obtained as
\begin{equation}\label{e3-5}
t=-\int_{\rho^{*}}^{\rho_{f}}\frac{d\rho}{\left(\frac{8\pi G}{3}\rho\right)^{1/2}\left(4\rho-\frac{16\xi}{3}
\rho^{5/4}\right)}\;,
\end{equation}
where $\rho^{*}$  represents an arbitrary initial finite value for density which lies in the intervals
between the fixed points. The negative sign in the back of integral relation, denotes a backward
in direction of time. We find that for the case of $\rho_{f_{1}}=0$, the integral (\ref{e3-5})
does not converge i.e. $|t|\rightarrow\infty$, while for the fixed point $\rho_{f_{2}}=(\frac{\epsilon
_{Pl}}{c})^{4}$  the integral (\ref{e3-5}) is not solvable
and its numerical values fail to converge. At this point, let us to draw the plot $O(\rho)-\rho$
(or $\dot{\rho}-\rho)$ in the figure 1, in order to get a qualitative analysis of the situation.
\begin{figure}[htp]
\begin{center}
\includegraphics{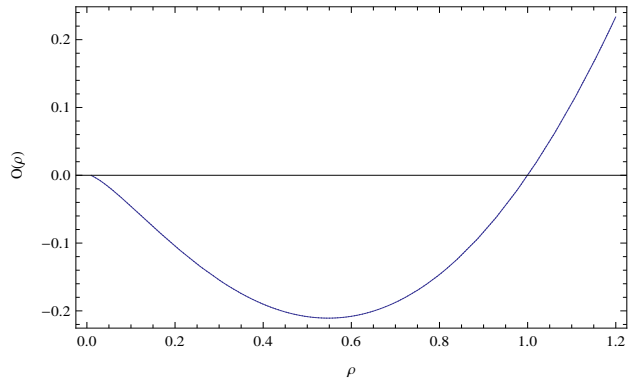} \vspace{4cm}
\end{center}
\caption{\small {The behavior of $O(\rho)$ (\ref{e3-3}) in terms of $\rho$  for flat
rainbow FRW Universe model. To simplify, we set $\epsilon_{pl}=c$ and $8\pi G=3$. Fixed
points are located in $\rho_{f_{1}}=0$ and $\rho_{f_{2}}=1$.}}
\label{fig:1}
\end{figure}
As it is seen from the figure 1, $O(\rho)$  has two fixed points $\rho_{f_{1}}=0$ and $\rho_{f_{2}}=1$.
In fact, any of these fixed points are equivalent to a de Sitter space since in these points we are
dealing with constant solution ($\dot{\rho}=0$). The first $\rho_{f_{1}}$ is a future fixed point
because for the case $\rho^{*}>0$ we have $O(\rho)<0$, while the second $\rho_{f_{2}}$ is an early (or past)
fixed point since for case of  $\rho^{*}<1$ and~ $\rho^{*}>1$ we have $O(\rho)<0$ and $O(\rho)>0$, respectively.
It is noteworthy that for distinction between the type of this singularities, we follow Ref. \cite{Adel}.
These fixed points classify the possible solutions into two classes: 1) a solution belongs to interval $\rho\in
[0,1]$, if $\rho_{f_{1}}<\rho^{*}<\rho_{f_{2}}$ and 2) a solution belongs to interval $\rho\in [1,\infty)
$, if $\rho^{*}>\rho_{f_{2}}$. Then, to get the fixed point $\rho_{f_{1}}$ by beginning from some
initial value as $\rho^{*}$  and also from $\rho_{f_{2}}$ to $\rho^{*}$, the time required is infinite.
In this interval, $O(\rho)$ is continuous and differentiable. Then, according to approach in \cite{Adel}
by pursuing the terminology of \cite{Strogatz}, one can say that the first solution is
free of physical singularity so that it interpolates steadily from $\rho_{f_{1}}$ to $\rho_{f_{2}}$. It is
clear that the first condition  already is not usable for the second solution, which represents a clear violation
of the second condition. On the other hand, for the case $\rho^{*}>\rho_{f_{2}}$, the function (\ref{e3-3}) is growing
quicker than a linear function. So, we conclude that the second solution is not free of a finite time
singularity. Because we are interested in studying initial singularity problem, so let us mention to
stability status of early fixed point $\rho_{f_{2}}$. A fixed point is stable if by putting a neighboring initial
value, the trajectory of the solution  remains always near
to the fixed point. Equivalently, a fixed point is unstable if for any point
at vicinity  of the fixed point, one can find some solution that starts near the fixed point but go away from it in a finite time.}
{The fixed point $\rho_{f_{2}}$ is an unstable point because  $\frac{dO(\rho)}{d\rho}|_{\rho=\rho_{f_{2}}}>0$} \footnote{{At a fixed point where $O(\rho_{f_{2}})=0$,
if $\frac{dO(\rho)}{d\rho}|_{\rho=\rho_{f_{2}}}>0$, we have increasing $O$   at $\rho_{f_{2}}$,
or equivalently $ f(\rho_{f_{2}}-\delta)<0<f(\rho_{f_{2}}+\delta)$ for all sufficiently small and positive $\delta$.
This means that if we start with initial value $\rho^{*}>\rho_{f_{2}}$ in the vicinity of $\rho_{f_{2}}$,
since $f(\rho^{*})>0$ then the trajectory of the ODE solution increases its value of $\rho_{f_{2}}$ and
moves away from the fixed point. Also if we start with $\rho^{*}<\rho_{f_{2}}$, but near to $\rho_{f_{2}}$,
again trajectory of the ODE solution moves away from the fixed point but  by decreasing its value
of $\rho_{f_{2}}$. Therefore, if $\frac{dO(\rho)}{d\rho}|_{\rho=\rho_{f_{2}}}>0$, we conclude that the fixed
point $\rho_{f_{2}}$ is unstable. Similarly, one can show
that if $\frac{dO(\rho)}{d\rho}|_{\rho=\rho_{f_{2}}}<0$,  the fixed point will be stable}.}.
{Then, although it takes  an  infinite time to get a fixed point, the issue of finite time singularities
avoidance remains unsolved and eventually this fixed point will collapse}. \\

{Now, we investigate the DGR theory form the
 emergent universe point of view in which there is no
 beginning of time and consequently, there is no big bang singularity in the early Universe. In this  scenario,
the Universe did not born from a Big-Bang singularity in a past finite time
and rather it possesses  an eternal Einstein static state. In this context,
the key point is the required conditions for the stability of the existing ESU,
in which we will explore in the following of the present paper}.
The ESU in the DGR scenario with rainbow functions varying with cosmological time can be obtained by the conditions
$\ddot a=\dot a=0$ through the equations (\ref{e2-a}) and (\ref{e2-b}) as
\begin{equation}\label{e3-6}
\frac{k}{a_{0}^{2}}=\frac{8\pi G\rho_{0}}{3}\;,
\end{equation}
and
\begin{equation}\label{e3-7}
-\frac{4\pi G\rho_{0}(1+3\omega)}{3f(\rho_{0})^{2}}=0\;,
\end{equation}
where $a_0$, $\rho_0$ and $\omega$  denote the scale factor, the energy density of the Einstein static Universe
and barotropic equation of state parameter ($p=\omega\rho$), respectively. Considering the positive energy condition
$\rho_0>0$ through the equation (\ref{e3-6}), it is seen that for the Einstein static Universe, the spatial curvature of
the Universe should be positive, $k>0$. Also, Eq.(\ref{e3-7}), shows that for the Einstein static Universe
in the framework of rainbow gravity, we need to have $\omega=-\frac{1}{3}$ or $f(\rho_{0})\rightarrow\infty$ corresponding to $\rho_{0}=\left(\frac{4}{3}\xi\right)^{-4}$.
The perturbations in the cosmic scale factor $a(t)$ and the energy density $\rho(t)$ can
be written as
\begin{eqnarray}\label{e3-8}
&&a(t)\rightarrow a_{0}(1+\delta a(t)),\nonumber\\
&&\rho(t)\rightarrow \rho_{0}(1+\delta \rho(t)).
\end{eqnarray}
Substituting (\ref{e3-8}) into the equation (\ref{e2-a}) after linearizing the perturbation terms,
we obtain  the following equation
\begin{equation}\label{e3-9}
-\frac{2k}{a_{0}^{2}}\delta a= \frac{8\pi G}{3}\rho_{0}\delta \rho\;.
\end{equation}
This indicates, with respect to the positiveness of $k$,  that the sign of  variation of the scale
factor must be opposite to the sign of variation of the matter density. We must also
apply the same method on
the equation (\ref{e2-b}), however before doing this let us introduce the following replacement due
to the perturbation in the rainbow function $f(\rho)$
\begin{equation}\label{e3-10}
f^{-2}(\rho)\rightarrow \left(f^{-2}(\rho_{0})-\frac{2}{3}\xi\rho_{0}^{\frac{1}{4}}
\delta\rho\right)\;,
\end{equation}
such that
\begin{equation}\label{e3-10a}
f^{-2}(\rho_{0})=1-\frac{8}{3}\xi \rho_{0}^{\frac{1}{4}}\;.
\end{equation}
Also, by combining the energy conservation equation (\ref{e2-c}) with the
first rainbow
modified FRW equation (\ref{e2-a}), we obtain
\begin{equation}\label{e3-11}
-\frac{\dot{a}}{a}\frac{\dot{f}}{f}=-\frac{\dot{a}}{a}\frac{df}{d\rho}.\frac{\dot
{\rho}}{f}=\frac{3(1+3\omega)}{4}\xi f^{-1}\rho^{\frac{1}{4}}
\left(\frac{8\pi G}{3}\rho-\frac{k}{a^{2}}\right).
\end{equation}
Then, after applying perturbation  (\ref{e3-8}), we have
\begin{equation}\label{e3-12}
-\frac{\dot{a}}{a}\frac{\dot{f}}{f}=\frac{3(1+3\omega)}{4}\xi\rho_{0}^{\frac{1}
{4}}(1+\frac{1}{4}\delta\rho)\left(f^{-1}(\rho_{0})-\frac{\xi}{3}\rho_{0}^{\frac
{1}{4}}\delta\rho\right)\left(\frac{8\pi G}{3}\rho_{0}(1+\delta\rho)
-\frac{k}{a_{0}^{2}}
(1-2\delta a)\right).
\end{equation}
By substituting (\ref{e3-6}) and (\ref{e3-9}) into the above expression, one finds that
$-\frac{\dot{a}}{a}\frac{\dot{f}}{f}=0$. It seems that this result is independent of
any specific form of rainbow function $f$. Therefore, it can be seen that the last
term of the second rainbow modified FRW equation (\ref{e2-b}) does not contribute
in derivation of stability conditions of Einstein static Universe in the framework
of DGR scenario. Now, putting perturbation equations (\ref{e3-8}) into (\ref{e2-b})
and using (\ref{e3-10}), we get
\begin{equation}\label{e3-13}
\delta\ddot{a}=-\frac{4\pi G}{3}(1+3\omega)\left(f^{-2}_{0}(\rho_{0})-\frac{2}{3}\xi
\rho_{0}^{\frac{1}{4}}\right)\rho_{0}\delta\rho\;,
\end{equation}
where inserting the equation (\ref{e3-9}) into (\ref{e3-13}) and neglecting the non-linear
perturbation terms, results in the following differential equation
\begin{equation}\label{e3-14}
\delta\ddot{a}-\frac{k}{a^{2}_{0}}(1+3\omega)\left(1-\frac{10}{3}\xi\rho_{0}^
{\frac{1}{4}}\right)\delta a=0.
\end{equation}
It is obvious that, for the cases $\xi\rightarrow0$, i.e. $\epsilon_{Pl}\rightarrow\infty$,
Eq.(\ref{e3-14}) will take the following form
\begin{equation}\label{e3-15}
\delta\ddot{a}-\frac{k}{a_{0}^{2}}(1+3\omega)\delta a=0\;,
\end{equation}
which refers to the oscillatory modes of ESU in the framework of standard GR for $\omega<-1/3$.
In order to have the general oscillating perturbation modes in the framework of the DGR scenario,
the following condition should be satisfied
\begin{equation}\label{e3-16}
-\frac{k}{a^{2}_{0}}(1+3\omega)\left(1-\frac{10}{3}\xi\rho_{0}^
{\frac{1}{4}}\right)>0\;,
\end{equation}
which results in the following solution for the equation (\ref{e3-14})
\begin{equation}\label{e3-17}
\delta a=\alpha_{1}e^{i\gamma_{0} t}+\alpha_{2}e^{-i\gamma_{0} t},
\end{equation}
where $\alpha_1$ and $\alpha_2$ are integration constants and $\gamma_{0}$ refers to
the frequency of oscillation around the stable ESU as
\begin{equation}\label{e3-18}
\gamma_{0}=\left(-\frac{k}{a^{2}_{0}}(1+3\omega)\left(1-\frac{10}{3}\xi\rho_{0}^
{\frac{1}{4}}\right)\right)^{\frac{1}{2}}
\end{equation}
Now, we can analysis the stability condition (\ref{e3-16}). Eqs.(\ref{e3-6}) and (\ref{e3-7})
will play the role of two fundamental constraints to achieve our goal. There are two possibility
in order to satisfy the equation (\ref{e3-7}); the first one is that $\omega=-\frac{1}{3}$ and the
second one is that the matter density of the ESU  be $\rho_{0}=\left(\frac{4}{3}\xi\right)^{-4}$.
If we set $\omega=-\frac{1}{3}$, then the condition (\ref{e3-16}) is
automatically violated and  we obtain $\delta\ddot{a}=0$ implying that there is no stable ESU
against the linear scalar perturbations in the form of relations (\ref{e3-8}). As it is clear from
Eq.(\ref{e3-15}), in the framework of the standard GR  for  $\omega=-\frac{1}{3}$,
there is no stable ESU. For the second possibility as $\rho_{0}=\left(\frac{4}{3}\xi\right)^{-4}$,
with respect to the positivity of the spatial curvature,
due to the positive energy condition through Equation (\ref{e3-6}), we need that the barotropic
EOS parameter satisfies $\omega>-1/3$.
Overall, one can find that
there is a stable ESU versus homogenous and linear scalar perturbations in context
of the gravity's rainbow scenario with choosing the rainbow function (\ref{e3-2}).
In this case, we need two  bounds which the first one in on EOS parameter as $\omega>-1/3$ and the second
one is on the
energy density of ESU as $\rho_{0}<\epsilon_{pl}^{4}$. There are no such
 similar results in the framework of the standard GR theory. Unlike GR, here there is a stable solution
 against small scalar perturbations for a closed universe filled by
the matter fields respecting the energy conditions. Also,
within standard GR theory and even in many of the modified gravity theories, there is no such
upper bounds on energy density of ESU $\rho_{0}$.
In the following, we want to investigate the stability conditions of Einstein static Universe against
the linear homogeneous scalar perturbations in the context of DGR, in terms of another rainbow function
proposed by Ling and et al in \cite{Hu}, which is evolving with cosmic  time as
\begin{equation}\label{e3-19}
f=\sqrt{1-l^{2}_{p}\bar{\epsilon}^{2}}.
\end{equation}
Considering previous arguments after placing the average energy $\bar{\epsilon}=\frac{4}{3}c
\rho^{\frac{1}{4}}$
\cite{Adel}, it can be rewritten in terms of energy density $\rho$ as
\begin{equation}\label{e3-20}
f(\rho)=\sqrt{1-\frac{16}{9}\chi\rho^{\frac{1}{2}}},
\end{equation}
where $\chi=l^{2}_{p} c^{2}$ is a constant. {Let us point out that rainbow
function (\ref{e3-20}) has application in the framework of black hole physics leading to
valuable phenomenological outcomes,  e.g. see \cite{Hu, mod}. Moreover, it also as previous rainbow
function has a theoretical trait. By reaching the energy density $\rho$ or $\bar{\epsilon}$ to
its largest value, $f(\rho)$ in (\ref{e3-20})  becomes zero which leads to vanishing spatial-like
components of the rainbow metric. This means that FRW metric modified by (\ref{e3-20}) or (\ref{e3-19}),
is a smooth and differentiable
metric but is not invertible. Moreover, by applying the rainbow functions (\ref{e3-19}) or (\ref{e3-20}),
we will be faced
with a degenerate metric at energy levels  close to the Planck energy scale.  Also, based on the fact that there
is no observational evidence for lightlike dimension in nature, there might be a suppression mechanism for appearance of lightlike dimension at
the quantum level of Universe \cite{Louko}}.\\
{As before,  we first examine the
Big-Bang singularity problem for such a choice of rainbow function within a flat rainbow FRW  model. In the same way, for the rainbow function
(\ref{e3-20}), we get
\begin{equation}\label{e3-21}
\dot{\rho}=-4\sqrt{\frac{8\pi G\rho^{3}}
{3-\frac{16}{3}\chi \rho\frac{1}{2}}},
\end{equation}
which has only one fixed point $\rho_{f_{1}}=0$ and is not bounded, see Figure 2. As it is  seen from the figure,
the function $O(\rho)$ begins from the fixed point $\rho_{f_{1}}$ and ends with a singularity. So we must
separately calculate the time needed from an infinite to a finite value  $\rho^{*}$ as well as the time needed from $\rho^{*}$ to the fixed point $\rho_{f_{1}}=0$, i.e,
\begin{equation}\label{e3-21a}
t= -\frac{1}{4}\int_{\infty} ^{\rho^{*}} d\rho \sqrt{\frac{3-\frac{16}{3}\chi \rho^{\frac{1}{2}}}{8\pi G\rho^{3}}},
\end{equation}
and
\begin{equation}\label{e3-21b}
t= -\frac{1}{4}\int_{\rho^{*}} ^{0} d\rho \sqrt{\frac{3-\frac{16}{3}\chi \rho^{\frac{1}{2}}}{8\pi G\rho^{3}}}.
\end{equation}
Here, we are dealing with a different situation compared to the previous one. We found that the integral (\ref{e3-21a})
converges which provide $\rho^*\geq\frac{81}{256\chi^2}$ while the integral (\ref{e3-21b})  does not converge anyway on the given interval. This means
that from $\rho^{*}$
to fixed point $\rho_{f_{1}}=0$  an infinite time is needed, i.e $|t|\rightarrow\infty,
$ while to
get an infinite to a finite value $\rho^{*}$ it takes a finite time. Because the rainbow FRW metric
includes a natural cutoff as the energy Planck,  the constraint obtained for integral (\ref{e3-21a}) should lie in the interval $\frac{81}{256\chi^2}\leq\rho^*\leq\frac{1}{\chi^2}$. Then,  the
solution derived from the rainbow function (\ref{e3-20}) for the flat rainbow Universe model can not be free
of finite time singularity.
\begin{figure}[htp]
\begin{center}
\includegraphics{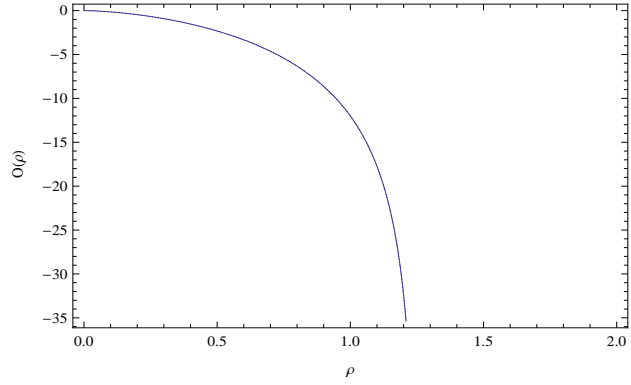} \vspace{4cm}
\end{center}
\caption{\small {{The behavior of $O(\rho)$ (\ref{e3-21}) in terms of $\rho$ for
the flat
rainbow FRW Universe model. To simplify, we set $8\pi G=3$ and $\chi=1/2$. }}}
 \label{fig:2}
\end{figure}
We conclude that even assuming that integral (\ref{e3-21a}) results in an infinite time,
the Big-Bang singularity issue is not canceled, yet.   Because according to the discussion given in the previous case,  one can
realize that $\rho_{f_{1}}$ is a future fixed point and is not a past one.} \\
{In the following, we study the
DGR theory from the emergent universe point of view and investigate 
the   required conditions for the stability of an ESU with respect to scalar perturbations, in the presence of rainbow function (\ref{e3-20}).}
By applying the scalar perturbation terms on the rainbow
function (\ref{e3-20}), it takes the form
\begin{equation}\label{e3-24}
f^{-2}(\rho)\rightarrow f^{-2}(\rho_{0})+\frac{8}{9}\chi\rho^{\frac{1}
{2}}_{0}\delta\rho\;,
\end{equation}
where
\begin{equation}\label{e3-25}
f^{-2}(\rho_{0})=1+\frac{16}{9}\chi\rho^{\frac{1}{2}}_{0}
\;.
\end{equation}
Now, inserting the equations (\ref{e3-24}), (\ref{e3-25}) and the perturbation equations
(\ref{e3-8}) into the second rainbow modified Friedmann equation (\ref{e2-b}), we have
\begin{equation}\label{e3-26}
\delta\ddot{a}=-\frac{4\pi G}{3}(1+3\omega)\left((\rho_{0}+1)f^{-2}(\rho_{0})
+\frac{8}{9}\chi\rho^{\frac{1}{2}}_{0}\right)\rho_{0}\delta\rho\;,
\end{equation}
where similar to the previous analysis, we neglected non-linear terms. Finally, by
applying the equations (\ref{e3-7}) and (\ref{e3-9}), the above differential equation takes
the following form
\begin{equation}\label{e3-27}
\delta\ddot{a}-\frac{k}{a^{2}_{0}}(1+3\omega)\left(1+\frac{8}{3}
\chi\rho^{\frac{1}{2}}_{0}\right)\delta a=0\;.
\end{equation}
As can be seen, for the limit $\chi\rightarrow0$, the above equation reduces to the oscillatory modes
(\ref{e3-15}). Then, in order to have a stable ESU in the framework of DGR scenario, the following
condition should be satisfied
\begin{equation}\label{e3-28}
-\frac{k}{a^{2}_{0}}(1+3\omega)\left(1+\frac{8}{3}
\chi\rho^{\frac{1}{2}}_{0}\right)>0\;,
\end{equation}
where the frequency of oscillatory modes $\gamma_{0}$ for the rainbow function (\ref{e3-20})
reads as
\begin{equation}\label{e3-29}
\gamma_{0}^{2}=-\frac{k}{a^{2}_{0}}(1+3\omega)\left(1+\frac{8}{3}
\chi\rho^{\frac{1}{2}}_{0}\right)\;.
\end{equation}
It is seen that for $\omega=-\frac{1}{3}$,  the condition (\ref{e3-28}) is
automatically  violated and we get $\delta\ddot{a}=0$ through the equation (\ref{e3-27}), which
indicates that there is no stable ESU against the linear scalar perturbations for this rainbow function.
For the second possibility as $\rho_{0}=\left(\frac{16}{9}\chi
\right)^{-2}$, with respect to the positivity of the spatial curvature
due to the positive energy condition through equation (\ref{e3-6}), we need that the barotropic EOS
parameter $\omega$ satisfies the condition $\omega<-1/3$. Therefore, here  we have a stable ESU
for the baratropic EOS parameter $\omega<-1/3$, denoting the phantom matter fields, with the bounded energy
density of ESU $\rho_{0}$ as $\rho_{0}< l_{pl}^{-4}$. As mentioned before, the existence of such constraint
on energy density of ESU may be absent in other modified gravity theories. Here, such a bound on initial
density $\rho_{0}$ is arising from
the energy dependent metric  in gravity's rainbow proposal which is one of the
common cut offs of  quantum gravity theories. Finally, it should
be mentioned that  choosing the rainbow function
(\ref{e3-20}) leads to the same result with GR so that a closed universe filled by usual non-relativistic
matter fields, is not stable against small linear scalar perturbations.
\subsection{In the Presence of Cosmological Constant $\Lambda$ }
In this subsection, we want to consider the possible modification by the cosmological constant
$\Lambda(\epsilon)$ and investigate its effects on stability condition of an ESU. In presence
of cosmological constant $\Lambda(\epsilon)$, the second modified FRW equation (\ref{e2-b}) remain
without change. But,  the first modified FRW equation (\ref{e2-a}) and the conservation law (\ref{e2-c}),
take the form of
\begin{equation}\label{e3-a1}
(\frac{\dot{a}}{a})^{2}+\frac{k}{a^{2}}\frac{1}{f^{2}}=\frac{8\pi G\rho}
{3}\frac{1}{f^{2}}+\frac{\Lambda(\epsilon)}{3f^{2}}\;,
\end{equation}
and
\begin{equation}\label{e3-a2}
\dot{\rho}+3\rho\frac{\dot{a}}{a}(1+\omega)=-\frac{\dot{\Lambda}(\epsilon)}{8\pi G}\;.
\end{equation}
One may consider another rainbow function as $h(\epsilon)$ so that $\Lambda(\epsilon)=h(\epsilon)^{2}\Lambda$
where $\Lambda$ is the usual cosmological constant. By this consideration,  the modified FRW equation (\ref{e3-a1}) will be
\begin{equation}\label{e3-a}
(\frac{\dot{a}}{a})^{2}+\frac{k}{a^{2}}\frac{1}{f^{2}}=\frac{8\pi G\rho}{3}
\frac{1}{f^{2}}+\frac{h(\epsilon)^{2}\Lambda}{3f^{2}}\;.
\end{equation}
By setting this new rainbow function as $h(\epsilon)^{2}=(1+\lambda
\rho)$, see second Ref. of \cite{Ling}, for the ESU described by $\dot{a}=\ddot{a}=0$, we  get the following equation from equation
(\ref{e3-a1}) as
\begin{equation}\label{e3-30}
\frac{k}{a_{0}^{2}}=\frac{8\pi G\rho_{0}}{3}+\frac{\Lambda}{3}(1+\lambda \rho_{0})\;,
\end{equation}
where $\lambda$ is a dimensional parameter. By keeping positive energy condition,
the above equation implies that positive spatial curvature $k>0$, for two cases $\Lambda>0$
and $\Lambda<0$, is guaranteed only under the following constraints, respectively
\begin{equation}\label{e3-301}
\lambda>-\frac{1}{\rho_{\Lambda}}-\frac{1}{\rho_{0}}\;,
\end{equation}
and
\begin{equation}\label{e3-302}
\lambda<\frac{1}{|\rho_{\Lambda}|}-\frac{1}{\rho_{0}}\;,
\end{equation}
so that $\rho_{\Lambda}\equiv\frac{\Lambda}{8\pi G}$ and $|\rho_{\Lambda}|\equiv\frac{|\Lambda|}{8\pi G}$.
With note to the our previous result on upper bounds on energy density  of ESU as $\rho_{0}<l_{pl}^{-4}$, the above
constraints reads as follows
\begin{equation}\label{e3-3022}
\lambda>-\frac{1}{\rho_{\Lambda}}-l_{pl}^{4}\;,
\end{equation}
and
\begin{equation}\label{e3-3002}
\lambda<\frac{1}{|\rho_{\Lambda}|}-l_{pl}^{4}\;.
\end{equation}
In DGR formalism of gravity, it is expected that parameter $\lambda$ to have the order of magnitude
$|\lambda|\sim l_{pl}^{4}$. As it seems, this value of $\lambda$ be satisfied both the above constraints.
Also, in the presence of cosmological constant, we recover Eq. (\ref{e3-7}) from the second modified FRW
equation (\ref{e2-b}).
Similar to the previous analysis,  by inserting equation (\ref{e3-8}) into the equation
(\ref{e3-a1}) and neglecting the non-linear perturbation terms, we get
\begin{equation}\label{e3-31}
-\frac{2k}{a_{0}^{2}}\delta a= \left(\frac{8\pi G}{3}+\frac{\Lambda}{3}\lambda \right)
\rho_{0}\delta \rho
\end{equation}
It is  seen that for the case of positive spatial curvature $k>0$, the sing of variation of the scale factor
$\delta a$ is in contrast to the variation of matter density $\delta \rho$, provided that
\begin{equation}\label{e3-32}
\lambda>-\frac{1}{\rho_{\Lambda}}\;,
\end{equation}
and
\begin{equation}\label{e3-322}
\lambda<\frac{1}{|\rho_{\Lambda}|}\;,
\end{equation}
where are related to the  cases $\Lambda>0$ and $\Lambda<0$, respectively. For the case of positive
cosmological constant, $\Lambda>0$,  by comparing the constraints (\ref{e3-301}) and
(\ref{e3-32}), one realizes that constraint (\ref{e3-301}) can be satisfied via constraint
(\ref{e3-32}) while its inverse is not true and so relation (\ref{e3-32}) will be a tighter constraint.
With the same reason, one can say that for the case of negative cosmological constant, $\Lambda<0$, constraint
(\ref{e3-302}) is more tighter than (\ref{e3-322}). Finally, by using  equation (\ref{e3-31})
and considering equation
(\ref{e3-8}) into the
second modified
FRW equation (\ref{e2-b}), after neglecting the non-linear perturbation terms, we arrive to
following differential equation
\begin{equation}\label{e3-33}
\delta\ddot{a}-\frac{k}{a^{2}_{0}}\left(\frac{1}{1+\lambda\rho_{\Lambda}}\right)
\left(1-\frac{10}{3}\xi\rho_{0}^
{\frac{1}{4}}\right)(1+3\omega)\delta a=0\;.
\end{equation}
By putting $\rho_{0}=\left(\frac{4}{3}\xi\right)^{-4}$, the above differential
equation takes the form of\begin{equation}\label{e3-34}
\delta\ddot{a}+\frac{3k}{2a^{2}_{0}}\left(\frac{1}{1+\lambda\rho_{\Lambda}}\right)
(1+3\omega)\delta a=0\;,
\end{equation}
In order to have a stable  ESU  against homogeneous and linear scalar
perturbation described by the oscillatory modes, we have two possibilities as
\begin{itemize}
\item For the case of $\Lambda>0$
\begin{equation}\label{e3-35}
\omega>-\frac{1}{3},\quad \lambda>-\frac{1}{\rho_{\Lambda}},\\
\quad \mbox{or} \quad \omega<-\frac{1}{3},\quad \lambda<-\frac{1}{\rho_{\Lambda}}.
\end{equation}
\item For the case of $\Lambda<0$
\end{itemize}
\begin{equation}\label{e3-36}
\omega>-\frac{1}{3},\quad\lambda<\frac{1}{|\rho_{\Lambda}|},\\
\quad \mbox{or} \quad \omega<-\frac{1}{3},\quad \lambda>\frac{1}{|\rho_{\Lambda}|}.
\end{equation}
Now, in the same way as above by using  the rainbow function (\ref{e3-2}), by
choosing the rainbow function
(\ref{e3-20}), we obtain the following differential equation
\begin{equation}\label{e3-37}
\delta\ddot{a}-\frac{k}{a^{2}_{0}}\left(\frac{1}{1+\lambda\rho_{\Lambda}}\right)
\left(1+\frac{8}{3}\chi\rho^{\frac{1}{2}}_{0}\right)(1+3\omega)\delta a=0\;,
\end{equation}
where by inserting $\rho_{0}=\left(\frac{16}{9}\chi\right)^{-2}$
, coming from constraint (\ref{e3-7}) for the rainbow function (\ref{e3-20}), it can be rewritten
as \begin{equation}\label{e3-38}
\delta\ddot{a}-\frac{5k}{2a^{2}_{0}}\left(\frac{1}{1+\lambda\rho_{\Lambda}}\right)
(1+3\omega)\delta a=0\;.
\end{equation}
This differential equation possesses the stable oscillatory modes  under the following possibilities\\
\begin{itemize}
\item The case of $\Lambda>0$
\begin{equation}\label{e3-39}
\omega>-\frac{1}{3},\quad \lambda<-\frac{1}{\rho_{\Lambda}},
\quad \mbox{or} \quad \omega<-\frac{1}{3},\quad \lambda>-\frac{1}{\rho_{\Lambda}}.
\end{equation}
\item The case of $\Lambda<0$
\end{itemize}
\begin{equation}\label{e3-40}
\omega>-\frac{1}{3},\quad \lambda>\frac{1}{|\rho_{\Lambda}|},
\quad \mbox{or} \quad \omega<-\frac{1}{3},\quad \lambda<\frac{1}{|\rho_{\Lambda}|}.
\end{equation}
 We note that since the cosmological constant does not appear in constraint equation (\ref{e3-7}), the energy
density of ESU is bounded as $\rho_{0}<\epsilon_{pl}^{4}$ or $ \rho_{0}< l_{pl}^{-4}$, for the both rainbow functions
(\ref{e3-2}) and (\ref{e3-20}). It is found that in order to have stable
solutions for an ESU in the presence of a cosmological constant, beside the
cutoff on energy density $\rho_{0}$,  the conditions  (\ref{e3-35}), (\ref{e3-36})
and (\ref{e3-39}), (\ref{e3-40}) for each of rainbow functions (\ref{e3-2}) and (\ref{e3-20}), should also be satisfied respectively. For the case of positive cosmological constant $\Lambda>0$, by comparing conditions mentioned in (\ref{e3-35})
and (\ref{e3-39}) with the constraint (\ref{e3-32}), one realize that with choice of rainbow functions (\ref{e3-2})
and (\ref{e3-20}), the solutions with $\omega>-\frac{1}{3},\quad \lambda>-\frac{1}{\rho_{\Lambda}}$
and $\omega<-\frac{1}{3},\quad\lambda>-\frac{1}{\rho_{\Lambda}}$ are allowed, respectively.
For the case of negative cosmological constant $\Lambda<0$, constraint (\ref{e3-302}), is more tighter
than (\ref{e3-322}). In fact, we recall that the  constraint (\ref{e3-302}) is important in the sense that with respect to physical energy
condition in quantum gravity regimes, ensures that $k>0$. So, it is clear that all the constraints on the parameter
$\lambda$ in (\ref{e3-36}) and (\ref{e3-40})  violate the constraint (\ref{e3-302}). Overall, this means that in the
presence of a quantum gravity modifications such as what is done by gravity's rainbow, with respect to positive energy condition and positive spatial
curvature $k>0,$  which is one of the basic assumptions in modern version of emergent universe scenario, for
the cosmological models possessing a negative cosmological constant,  there is no stability for an ESU.
\section{Einstein Static Universe, Vector and Tensor Perturbations}
In the cosmological context, the vector perturbations of a perfect fluid having energy
density $\rho$ and barotropic pressure $p=\omega\rho$ are governed by the co-moving
dimensionless {\it vorticity} defined as ${\varpi}_a=a{\varpi}$. The vorticity modes
satisfy the following propagation equation \cite{tensor}
\begin{equation}\label{v-1}
\dot{\varpi}_{\kappa}+(1-3c_s^2)H{\varpi}_{\kappa}=0,
\end{equation}
where $c_s^2=dp/d\rho$ and $H$  are the sound speed and the Hubble parameter, respectively.
This equation is valid in our treatment of Einstein static Universe in the framework of the
rainbow gravity through the modified Friedmann equations (\ref{e2-a}) and (\ref{e2-b}).
For the Einstein static Universe with $H=0$, the equation (\ref{v-1}) reduces to
\begin{equation}\label{v-2}
\dot{\varpi}_{\kappa}=0,
\end{equation}
where indicates that the initial vector perturbations remain frozen and consequently we have
neutral stability against the vector perturbations. Tensor perturbations, namely gravitational-wave
perturbations, of a perfect fluid is described by the co-moving dimensionless transverse-traceless
shear tensor $\Sigma_{ab} =a\sigma_{ab}$, whose modes satisfy the following equation
\begin{equation}\label{v-3}
\ddot\Sigma_{\kappa}+3H\dot\Sigma_{\kappa}+\left[\frac{\mathcal{K}^2}{a^2}
+\frac{2k}{a^2}-\frac{(1+3\omega)\rho-2\Lambda(\epsilon)}{3}\right]\Sigma_{\kappa}=0,
\end{equation}
where $\mathcal{K}$ is the co-moving index ($D^2\rightarrow -\mathcal{K}^2/a^2$ in which $D^2$ is
the covariant spatial Laplacian)\cite{tensor}. For the Einstein static Universe
$(H=0)$, this equation  reduces to
\begin{equation}\label{v-4}
\ddot\Sigma_{\kappa}+\left[\frac{\mathcal{K}^{2}}{ a_0^2} +\frac{2k}{a_{0}^2}-\frac{(1+3\omega)\rho_{0}-
2\Lambda(1+\lambda \rho_0)}{3}\right]\Sigma_{\kappa}=0.
\end{equation}
{Note that in the general tensor perturbation
equation (\ref{v-3}), the Hubble parameter $H$  contains the
rainbow  function $f$ through equation (\ref{e2-a}). But, because   for the Einstein
static universe we have $H=0$, then the rainbow  function $f$ did not
appear in the equation (\ref{v-4}).
} Then,  in order to have stable modes against the tensor perturbations, the following inequality should be satisfied
\begin{equation}
\frac{\mathcal{K}^{2}}{ a_0^2} +\frac{2k}{a_{0}^2}>\frac{(1+3\omega)\rho_{0}-
2\Lambda(1+\lambda \rho_0)}{3}.
\end{equation}
 Where, for a closed universe with $k=1$, considering the eigenvalue spectra  $\mathcal{K}^2=n(n+2)$ with $n=1,2,3,...$ \cite{Har}, will takes the form
of
\begin{equation}
\frac{n^2 +2n +2}{a_{0}^2}>\frac{(1+3\omega)\rho_{0}-
2\Lambda(1+\lambda \rho_0)}{3}.
\end{equation}
This inequality gives a restriction on  the scale factor of an ESU in terms of its
matter density and background cosmological constant.
Even though the existence of open and flat solutions  are possible in some modified gravity models, in the present modified gravity model  it is forbidden by  regarding the weak energy condition. Therefore, the stability
analysis in the present model is restricted to the physically viable
closed cosmological model.
For the the eigenvalue spectra  $\mathcal{K}^2$ for open and flat models,
one is referred to  \cite{Har}.
\section{Summery and Conclusion}
According to a modern version of emergent cosmological scenario
proposed by Ellis and Maartens \cite{EM}, the early Universe before passing to the inflationary
phase, has experienced an eternal Einstein static state rather than a Big-Bang singularity.
More exactly, it refers to a static closed space in the asymptotic past (before
entering the Universe to the period of inflation) known as ``Einstein static Universe'' (ESU).
The initial conditions as the quantum gravity effects in the early high energy Universe, will
influence the stability of this  static state . For this reason,
in the present work, we have examined the effects of an effective approach to quantum gravity proposed by
Magueijo and Smolin \cite{Magueijo}, known as "gravity's rainbow" or "doubly
general relativity (DGR)", on the stability of the Einstein
static state against the linear homogeneous scalar, vector and tensor perturbations. In order to following
our aim in the framework of DGR, we have needed to introduce appropriate
rainbow functions. Then, we have considered two appropriate well known rainbow functions (\ref{e3-2}) and (\ref{e3-20}).
{First, by following the phase space mechanism represented in \cite{Adel},  we have
investigated the fixed points belonging to a flat FRW model which is modified by the presence of the  rainbow function (\ref{e3-2}). Indeed, these
fixed points are de Sitter space solutions of typical flat rainbow FRW model. Our results indicate that
although the needed time to get a fixed point is infinite, it may not lead to the elimination of the initial finite time
singularities (Big-Bang). This is because of the fact that the fixed points corresponding to past singularities (Big-Bang) are unstable
and  may be collapse. Then, we studied the
DGR theory in the context of  the emergent universe scenario   and tried
to find its stable Einstein static
 universe with the required conditions}.
In order to find a stable ESU, as the first result, we realized that similar to standard GR scenario and independent of the choice
of rainbow function, the positive spatial curvature $k>0$, is the only option respecting to positive
energy condition $\rho_{0}>0$ for an ESU in the framework of the DGR theory.  For the case of zero
cosmological constant $\Lambda=0$, we found that in order to achieve a stable ESU against homogeneous
and linear scalar perturbations, we need a conditions on the energy density of ESU and the barotropic
equation of state as $\rho_{0}=\left(\frac{4}{3}\xi\right)^{-4}$ and $\omega>-\frac{1}{3}$, respectively.
{Similarly, for the rainbow function (\ref{e3-20}), we also verified that
Big-Bang singularity, can still exists. Unlike the previous case, the time needed to get the fixed point
is not infinite in this case. Furthermore, the only fixed point which is
revealed in the presence of rainbow function (\ref{e3-20}),
is a future fixed point and is not a past fixed point. Then, by looking at
the
DGR theory from the emergent universe point of view,  we studied  the stability of the Einstein static
 universe and its required conditions}.  For this case, a stable solution for ESU is guaranteed under these conditions
$\rho_{0}=\left(\frac{16}{9}
\chi\right)^{-2}$ and $\omega<-\frac{1}{3}$ describing the exotic matter fields. As it is
seen, stability conditions $\rho_{0}=\left(\frac{4}{3}\xi\right)^{-4}$ and $\rho_{0}=\left(\frac{16}{9}\chi\right)
^{-2}$ are direct result from the idea of an energy dependent metric in the framework of gravity's rainbow scenario.
These results
are equivalent to an explicit cutoff on the energy density of an ESU as $\rho_{0}<\epsilon_{pl}^{4}$ and $\rho_{0}<l_{pl}^{4}$,
respectively. Given that, ESU point out a initial static state (or static closed space) of universe before
getting into inflationary phase, so the existence of such explicit cutoff (or upper bounds) on $\rho_{0}$
could be interpreted as a result of initial dominate quantum gravity effects such as ``gravity's rainbow''.
In the  following of our analysis, we take an energy dependent cosmological constant via the introduction
of new rainbow function as $h(\epsilon)^{2}=(1+\lambda\rho)$. It is seen that the  positive spatial curvature,
$k>0,$ dictates an opposite sing for $\delta a$ relative to energy density $\delta \rho$ through the equation
(\ref{e3-31}). We find that in order to have stable
solutions for an ESU,beside the
cutoff on energy density $\rho_{0}$,  the conditions  (\ref{e3-35}), (\ref{e3-36})
and (\ref{e3-39}), (\ref{e3-40}) for each of rainbow functions (\ref{e3-2}) and (\ref{e3-20}), should also be satisfied respectively.
In particular, for the case of positive cosmological constant $\Lambda>0$, by comparing conditions mentioned in (\ref{e3-35})
and (\ref{e3-39}) with the constraint (\ref{e3-32}), one realize that with choice of rainbow functions (\ref{e3-2})
and (\ref{e3-20}), the solutions with $\omega>-\frac{1}{3},\quad \lambda>-\frac{1}{\rho_{\Lambda}}$
and $\omega<-\frac{1}{3},\quad\lambda>-\frac{1}{\rho_{\Lambda}}$ are allowed, respectively.
For the case of negative cosmological constant $\Lambda<0$, constraint (\ref{e3-302}), is more tighter
than (\ref{e3-322}). In fact, we recall that the  constraint (\ref{e3-302}) is important in the sense that with respect to physical energy
condition in quantum gravity regimes, ensures that $k>0$. So, it is clear that all the constraints on the parameter
$\lambda$ in (\ref{e3-36}) and (\ref{e3-40})  violate the constraint (\ref{e3-302}). Overall, this means that in the
presence of a quantum gravity modifications such as what is done by gravity's rainbow, with respect to positive energy condition and positive spatial
curvature $k>0,$  which is one of the basic assumptions in modern version of emergent universe scenario, for
cosmological models with a negative cosmological constant,  there is no stability for an ESU.
Finally, we investigated the the stability of an ESU
in the framework of DGR versus vector and tensor perturbations.  It is found
that  there is a
neutral stability against the vector perturbations. In order to have the
stability against the tensor
perturbations,  the scale factor of an ESU is restricted by  its
matter density and background cosmological constant.
\section*{Acknowledgment}
This work has been supported financially by Research Institute
for Astronomy and Astrophysics of Maragha (RIAAM) under research project
NO.1/3720-3.

\end{document}